\documentclass[a4paper,11pt]{article}
\pdfoutput=1 

\usepackage{jheppub} 

\usepackage[T1]{fontenc} 
\usepackage{cancel}

\title{BaryoGEN, a Monte Carlo Generator for Sphaleron-Like Transitions in Proton-Proton Collisions}

\author[1]{Cameron Bravo\note{Corresponding author.}}
\author{and Jay Hauser}

\affiliation{Department of Physics and Astronomy, University of California, Los Angeles, CA 90095-1547, USA}

\emailAdd{cbravo135@ucla.edu}

\abstract{
    Sphaleron and instanton solutions of the Standard Model provide violation of 
    baryon and lepton numbers and could lead to spectacular events at the LHC or future colliders. 
    Certain models of new physics can also lead to sphaleron-like vacuum transitions.
    This nonperturbative physics could be relevant to the generation of 
    the matter-antimatter asymmetry of the universe.
    We have developed BaryoGEN, an event generator that facilitates the exploration of
    sphaleron-like transitions in proton-proton
    collisions with minimal assumptions.
    BaryoGEN outputs standard Les Houches Event files 
    that can be processed by PYTHIA, and the code is publicly available. 
    We also discuss various approaches to experimental searches for such
    transitions in proton-proton collisions.
}

\begin{document} 
\maketitle
\flushbottom

\section{Introduction}
The class of solutions of gauge field theories to which the sphaleron 
belongs were first proposed in 1976 by 't Hooft \cite{tHooft:76}. 
These solutions are nonperturbative, so the cross-sections for processes mediated by the sphaleron cannot be calculated
perturbatively, e.g.\ by using Feynman diagrams. The solutions are high-energy but are unstable and decay
immediately.

The electroweak (EW) sphaleron was first described in 1984 \cite{Manton:84}. It is also
a critical piece of a leading cosmological model for the generation of the matter-antimatter
asymmetry in the universe known as 
Electroweak Baryogenesis \cite{Trodden:99} as well as thermal Leptogenesis \cite{leptogen}. 
The critical aspect of the sphaleron used in this model is the 
violation of baryon number ($B$) and lepton number ($L$) with conservation of $B-L$. 
Classically there exist 12 globally conserved U(1) currents, corresponding
to the conservation of baryon and lepton numbers.
Under reasonable approximations the $(B+L)$-violating
phenomenology is summarized as follows \cite{TEU_Kolb}.
An anomaly breaks the conservation of the U(1) currents \cite{tHooft:76}:
\begin{equation}
    \partial_\mu J^\mu = \frac{g^2}{16 \pi^2} \text{Tr}[F_{\mu\nu} \tilde{F}^{\mu\nu}],
\end{equation}
where $F_{\mu\nu}$ is the SU(2)$_L$ field strength tensor. The integral of this term can be non-zero:
\begin{equation}
    N_{CS} = \frac{g^2}{16 \pi^2} \int d^4x \text{Tr}[F_{\mu\nu} \tilde{F}^{\mu\nu}].
\end{equation}
This integral is known as a Chern-Simons \cite{CSnumber} or winding number.
The parameter $N_{CS}$ is a continuous characteristic of the gauge field with a periodic vacuum potential. The local minima of
this potential are at integer values of $N_{CS}$ while the sphaleron solutions exist at the local maxima, where
$N_{CS}$ is a half-integer. Sphaleron-like transitions are characterized by the 
$\Delta N_{CS}$ of the vacuum transition
they mediate. The anomaly exists equally for each fermion doublet. 
This means that, due to the vacuum transition, $L$ changes by $\Delta N_{CS}$ for each of the three
lepton doublets, since each lepton has $L=1$ or $L=-1$, while 
$B$ changes by $3 \Delta N_{CS}$ because
each quark doublet has $B=1/3$
and there are three colors and three families.
This results in two important relations that are essential to the phenomenology of sphalerons \cite{TEU_Kolb}:
\begin{equation}
    \begin{aligned}
            \Delta(B+L) &= 6 \Delta N_{CS}  \\
            \Delta(B-L) &= 0.
    \end{aligned}
\end{equation}
The sphaleron cannot exist without enough energy to overcome the periodic potential in $N_{CS}$, which can be described given all of the electroweak couplings. 
Since the discovery and mass measurement of the Higgs boson \cite{CMShiggs,ATLAShiggs},
all of these couplings have been measured, enabling the calculation of the minimum energy 
required for sphaleron-like transitions as $E_{Sph} = 9$~TeV \cite{Manton:84,TyeWong}. 


In general, as described below, sphaleron-like transitions in proton-proton 
scattering will produce spectacular events containing large numbers of energetic particles
that are similar to signatures expected from hypothetical microscopic black holes.
Tye and Wong \cite{TyeWong} argue that
the "pre-exponential factor" governing the rate of these transitions might not
be much less than one. This argument is currently under debate and has been
criticized in \cite{TWC1,TWC2} and followed up by Tye and Wong \cite{TWF1,TWF2}.
Ellis and Sakurai \cite{Ellis2016} present
a study of such transitions in proton-proton collisions by reinterpreting
the results of a search for microscopic black holes by ATLAS \cite{ATLASbh}.
Beyond Standard Model (BSM) fermions can also lead to significant rates \cite{Sakurai18}.
In this paper, we describe BaryoGEN, a new and more general event generator for 
sphaleron-like transitions in proton-proton collisions. 
BaryoGEN interfaces to general-purpose tools such as PYTHIA \cite{PYTHIA} via 
Les Houches Event (LHE) files \cite{LHE}, 
and the code has been made public \cite{BaryoGEN}.
BaryoGEN is relevant in general for any model violating baryon and lepton-number 
via a $\Delta N_{CS} = \pm 1$ vacuum transition due to a chiral anomaly, 
even for BSM searches at LHC energies.

\section{Physics Content}
We introduce the phenomenology used to determine the
final-state particle content of minimal sphaleron-like transitions.
We ignore the possibility of having large multiplicities of gauge bosons as in \cite{RINGWALD19901,ESPINOSA1990310}
since such transitions would be even easier to identify in experimental data than the minimal transitions. 

\subsection{Fermionic Content of Transitions}
Sphaleron-like transitions in proton collisions mediate processes such as:
$$u + u \rightarrow e^+ \mu^+ \tau^+ \bar{t} \bar{t} \bar{t} \bar{c} \bar{c} \bar{c} \bar{d} \bar{d} \bar{d} + u u + X \quad (\Delta N_{CS} = -1)$$
or
$$u + u \rightarrow \nu_e \mu^- \nu_\tau t b b  c s c  d d u + u u + X \quad (\Delta N_{CS} = +1),$$
where X is a set of particles that has $B = L = 0$.
BaryoGEN produces events resulting from both $\Delta N_{CS} = +1$ and $\Delta N_{CS} = -1$ transitions.
In all generality, the fermion configurations for the $(B+L)$-violating part of the transitions can be written as:
\begin{equation}
\epsilon_{ab} \epsilon_{cd}  \epsilon_{ef} \epsilon_{gh} \epsilon_{ij}  \epsilon_{kl} D^a_\alpha D^b_\beta  D^c_\gamma D^d_\delta D^e_\zeta  D^f_\mu D^g_\theta D^h_\eta  D^i_\iota D^j_\nu D^k_\kappa  D^l_\lambda,
\end{equation}
where each D represents a fermion doublet that is chosen uniquely from the set of 12 doublets by the lower indices, while the upper indices are the SU(2) indices. 
The six epsilons guarantee conservation of the electroweak charges. To determine the number of quantum-mechanically unique
final states, the first step is to assign a different lower index number 
to each of the 12 doublets in a sphaleron-like transition. There is
a single lepton doublet and three quark doublets (one for each color) from each generation. The epsilons then require all of
the doublets to be paired. The number of ways to pair 12 unique objects is:
$$\frac{1}{6!} {\displaystyle\prod_{n=1}^{6} {\binom{2n}{2}}} = 10395.$$
There is an additional factor of two for each epsilon because the SU(2) indices can be chosen in two ways, and one final factor of two
for the sign of $N_{CS}$ of the transition. This makes a total of $1,330,560$ quantum mechanically unique $\Delta N_{CS} = \pm 1$ transitions.

Absent any argument to favor some of the $(B+L)$-violating fermion configurations over others, 
BaryoGEN uses the least restrictive assumption that they occur equally.
We note that many of these final states will appear nearly identical in collider experiments,
with only 32 phenomenologically distinct final states of sphaleron-like transitions,
coming from eight distinctive lepton configurations and four distinctive quark
configurations.
The eight distinctive lepton configurations are counted as follows:
there is one configuration with $e, \mu$, and $\tau$; three configurations with
two different generations of charged leptons and one neutrino; three configurations with one
charged lepton and two neutrinos; and one configuration with three neutrinos. 
The four distinctive quark configurations are counted as follows:
the three third-generation quark doublets lead to four rather distinguishable final states; 
$ttt$, $ttb$, $tbb$, and $bbb$; while the other six quark doublets
result in first or second generation "light quark" jets that are nearly indistinguishable.
Additionally, the number of light quark jets may be affected by consideration 
of the incoming partons, as described below.

\subsection{Incoming Partons and Cancellations}
When producing the set of outgoing fermions the two incoming partons are added to 
the generated set of 12 outgoing fermions. If this addition results in any 
quark-antiquark pairs, they are removed. 
This generator therefore cancels zero, one, or two quarks, 
whereas the treatment in HERBVI \cite{HERBVI} only considers
the case of $\Delta N_{CS} = -1$ with exactly two cancellations.
Ellis and Sakurai \cite{Ellis2016} only 
allowed for zero or two cancellations for $\Delta N_{CS} = $ 1 or -1, respectively.
The prior treatments implicitly assume negligible sea quark interactions, 
which might seem a priori reasonable for $\sqrt{s} =$~14~TeV, 
but less so at higher energies such as 100~TeV.

\begin{figure}
    \centering
    \includegraphics[width=.45\textwidth]{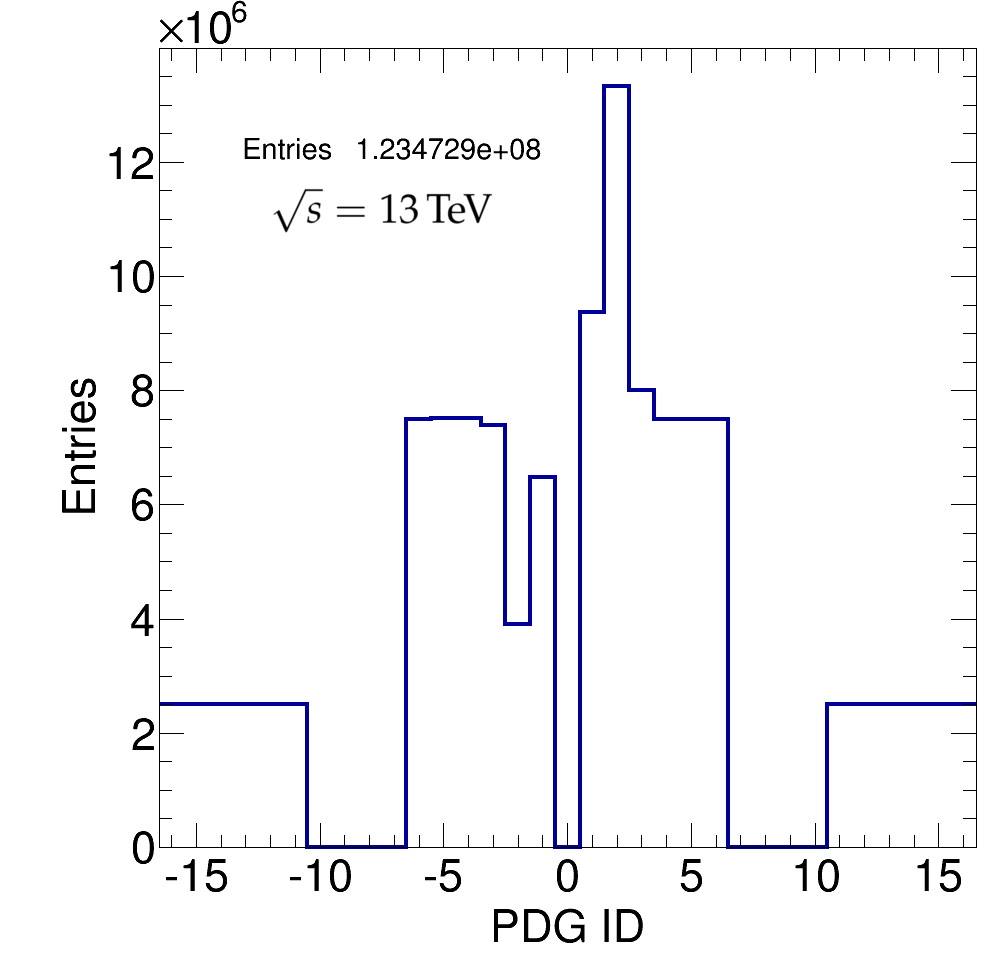}
    \caption{\label{fig:outIDs} The number of each of the fermion types (PDG IDs \cite{PDG}) coming directly from $10^7$ sphaleron-like transitions and CT10 PDFs.}
\end{figure}

Zero quark cancellations yield a 14-fermion final state, including three leptons, 
three third-generation quarks, and eight light quark jets.
With one cancellation there are 12 outgoing fermions, 
of which six are light quark jets, and
with two cancellations there are 10 outgoing fermions, 
of which only four are light quark jets.
We make a simple ansatz that
$N_{CS} = -1$ and $N_{CS} = +1$ transitions are equally likely.
We also use the CT10~\cite{CT10} Parton Distribution
Function (PDF) set throughout this article.
Under these assumptions, the relative rates of these three cases 
are shown in Table~\ref{tab:Nlhe}. 
It is observed that double cancellations are rare when $\Delta N_{CS}=+1$, and
zero cancellations are uncommon when $\Delta N_{CS}=-1$, but
single cancellations are actually quite common for $\Delta N_{CS}=-1$.
The distribution of the PDG~ID 
\cite{PDG} of the fermions directly from the transition are shown in Figure \ref{fig:outIDs}. 
The excess for up quarks (PDG ID $+2$) and deficit for up antiquarks (PDG~ID~$-2$) is due to the up quarks being the most common incoming partons, which are thus the most probable 
to be added as an extra outgoing fermion in the $\Delta N_{CS}=+1$ case,
or to cancel an outgoing up antiquark in the $\Delta N_{CS}=-1$ case.

It may be noted that the values in Table~\ref{tab:Nlhe} depend on the relative
levels of sea and valence quarks at large $x$ and at extremely high values of $q^2$ 
not yet directly probed by experiment; in particular a large variation is 
seen in NNPDF3.0~\cite{NNPDF} with respect to other choices of PDF sets.

\begin{table}[htbp]
    \centering
    \begin{tabular}{c|c|c}
           Multiplicity &  $N_{CS} = -1$ & $N_{CS} = +1$ \\
        \hline
                  10 &  27.9\%  &  0.0\% \\
                  12 &  19.1\%  &  0.5\% \\
                  14 &   3.0\%  & 49.5\% \\
    \end{tabular}
    \caption{\label{tab:Nlhe} The fractions of each fermion multiplicity of sphaleron-like transitions, with $\sqrt{s}=13$ TeV, and equal probabilities for
$N_{CS}=-1$ and $N_{CS}=+1$, from a generated sample of $10^7$ events and CT10 PDFs.}
\end{table}

\subsection{Color Flow}

In order to use PYTHIA to decay and hadronize the hard processes made by this generator, color flow lines must be drawn and enumerated.
PYTHIA can only handle baryon number violation of one at a given vertex, and since sphaleron-like
transitions violate baryon number by 3, the transition is factorized by introducing fake mediators in the LHE file. These mediators are defined for convenience with PDG IDs of 
1000022 ($\tilde\chi^0_1$), 
1006213 ($R^+$), and 1006223 ($R^{++}$). 
Note that HERWIG~\cite{HERWIG} must be configured to recognize these particle IDs
in order to use it with the output LHE files created by this generator.

\begin{figure}
    \centering
    \includegraphics[scale=0.30]{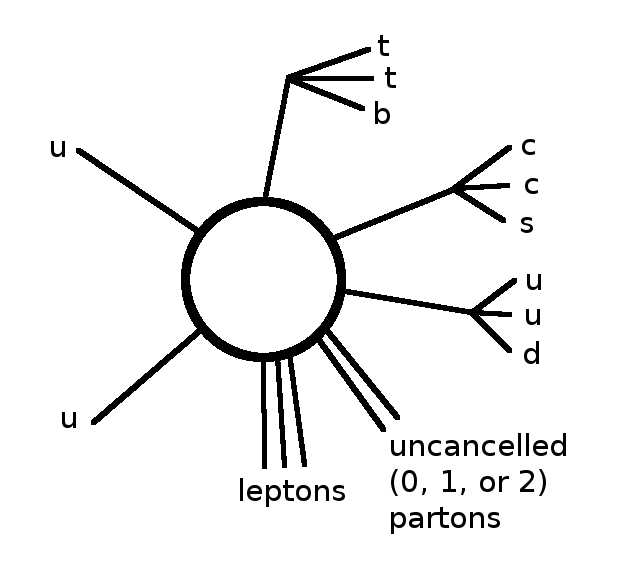}
    \caption{\label{fig:colFlowEx} An example of a sphaleron-like transition event
        starting from an initial state of two up quarks. The diagram is a representation of the
        event as written to the LHE file in order to allow correct determination of color flow by
        PYTHIA for decay and hadronization.}
\end{figure}

The outgoing quarks, after cancellations, are first grouped together by generation. 
An example of the graph represented by an event in the outgoing
LHE file is shown in Figure \ref{fig:colFlowEx}. This particular transition vertex has two incoming quarks, the outgoing leptons, the outgoing uncanceled incoming quarks, and
a fake mediator for each generation of outgoing quarks. 

Each generation of outgoing quarks must be a color singlet, otherwise it is possible to
get two identical fermion states, which is not allowed by the Pauli exclusion principle. 
Each generation of quarks is given a fake mediator to ensure that PYTHIA builds this set of quarks into a color singlet.
In the case of quark cancellation, the mediator will share a color line with the canceled quark. If it is ever the case that two
quarks are canceled from the same generation, the mediator is retracted and the only remaining quark from that particular generation comes directly off the central vertex.

After the fermion content of the final state is determined, the fermions are given random momenta according to phase space using ROOT \cite{ROOT} class {\bf TGenPhaseSpace}, 
which in turn invokes the GENBOD function from CERNLIB.

\subsection{Simulation Results} \label{simresults}

Simulations were conducted at the center of mass energies of 13, 14, 28, and 33~TeV. 
The corresponding cross-sections for a pre-exponential factor of one for these beam
 energies are, respectively,
$10.05$, $50.72$, $111,958$, and $285,053$ $\text{fb}^{-1}$.
A total of $10^5$ events were produced and then processed with PYTHIA at each energy.
These simulations were made with the nominal $E_{Sph} = 9$~TeV and, as
previously mentioned, CT10 PDFs and the ansatz that
$N_{CS} = -1$ and $N_{CS} = +1$ transitions are equally likely.
We count the number $N$ of quarks and charged leptons
with transverse momentum $p_T >$~20~GeV and pseudo-rapidity $|\eta| < 2.5$, 
so that the results can be easily compared with those in \cite{Ellis2016}.

\begin{figure}
    \centering
    \includegraphics[scale=0.20]{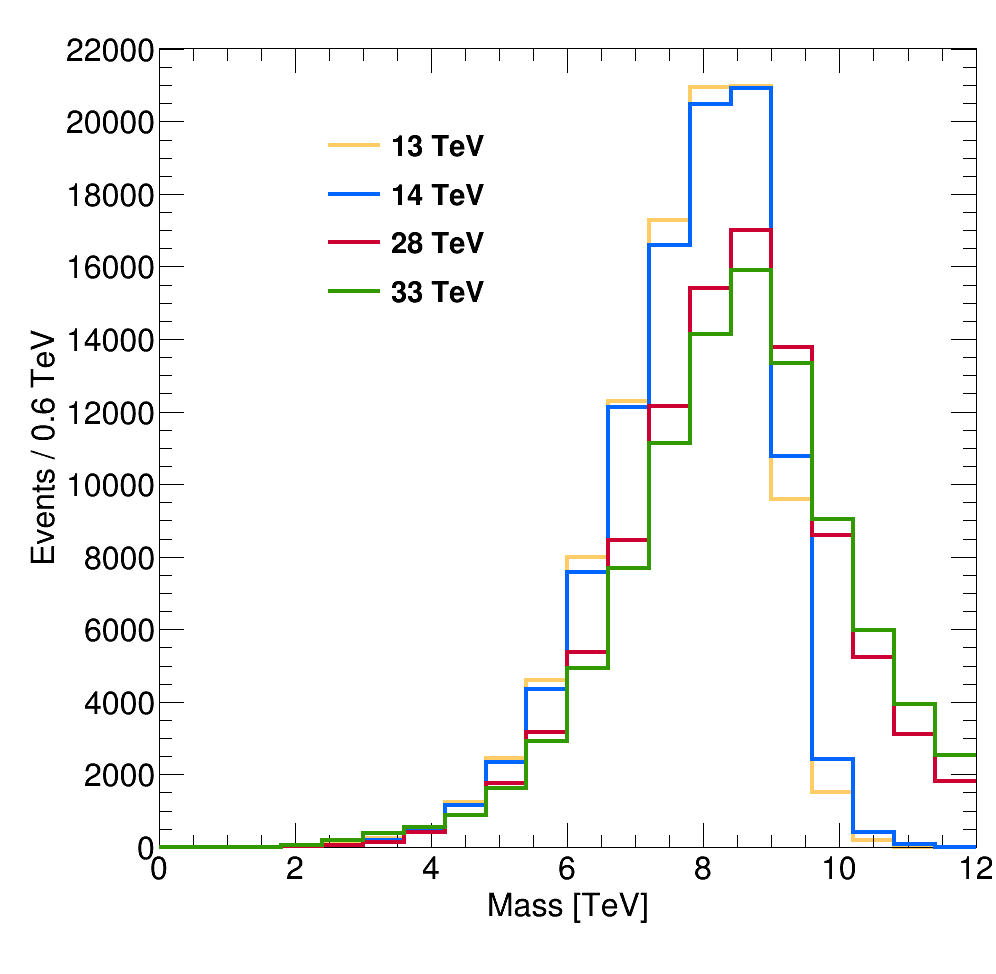}
    \caption{\label{fig:massSph} Invariant-mass distributions of observable 
        final-state particles in simulated LHC collisions at 13, 14, 28, and 33 TeV. 
        These simulations are made with the nominal $E_{Sph}=9$~TeV.}
\end{figure}

\begin{figure}
    \centering
    \includegraphics[scale=0.2]{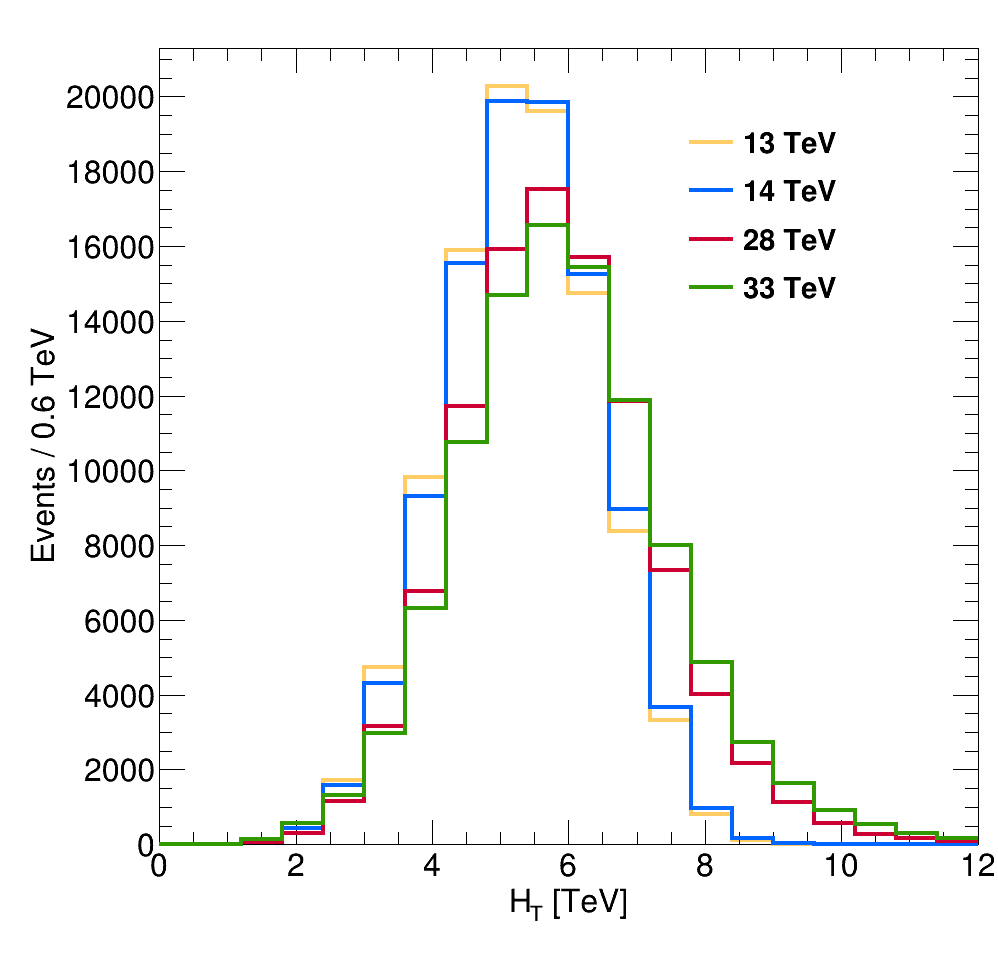}
    \includegraphics[scale=0.2]{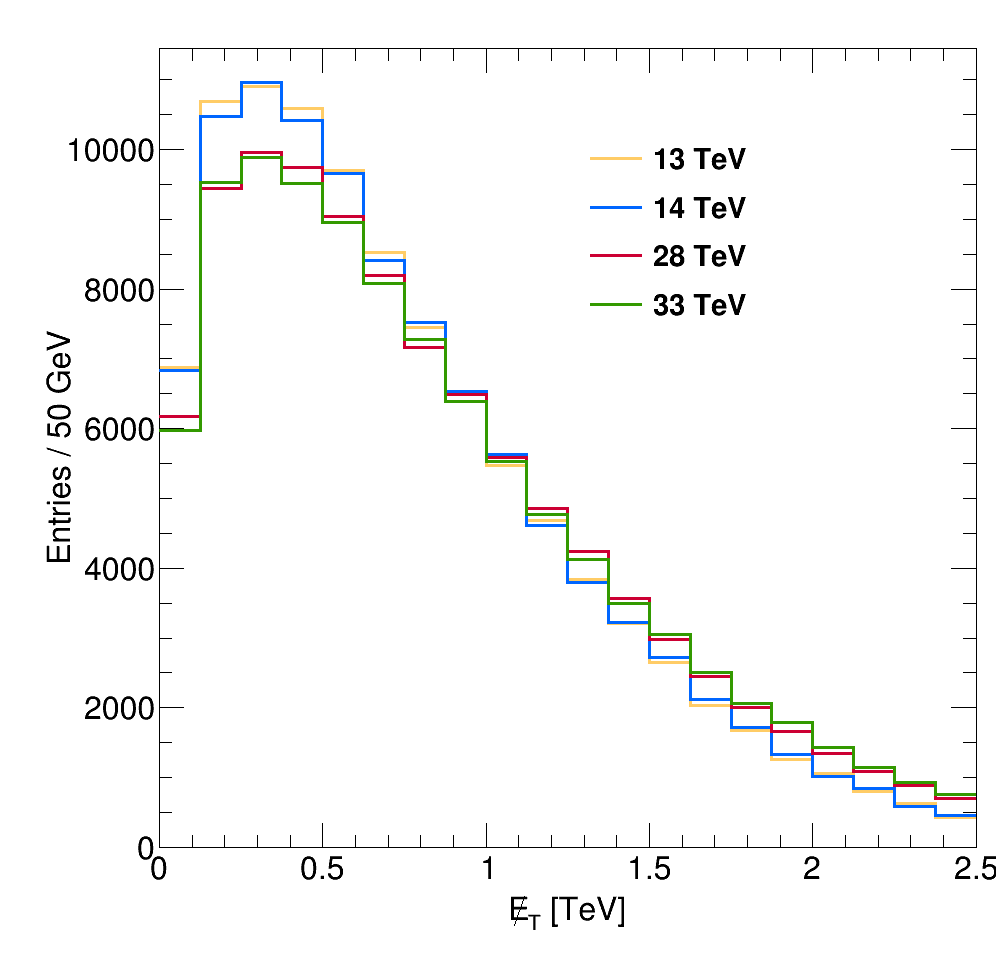}
    \caption{\label{fig:kinDist} Left Panel: Distributions of the scalar sum of 
        $p_T^{jet}$ of events 
        in simulated LHC collisions at 13, 14, 28, and 33 TeV. 
        Right Panel: Distributions of $\cancel{E}_T$
        in simulated LHC collisions at all four energies.
        These simulations are made with the nominal $E_{Sph}=9$~TeV.}
\end{figure}

Distributions of observable invariant mass are presented in 
Figure~\ref{fig:massSph}, comparable to Figure~3 (left) of \cite{Ellis2016}.
The total transverse energy of all jets $H_T$, and 
the missing transverse energy $\cancel{E}_T$, are shown in Figure \ref{fig:kinDist}, comparable
to Figure~5 (top left and bottom left, respectively) of \cite{Ellis2016}. 

\begin{figure}
    \centering
    \includegraphics[width=.45\textwidth]{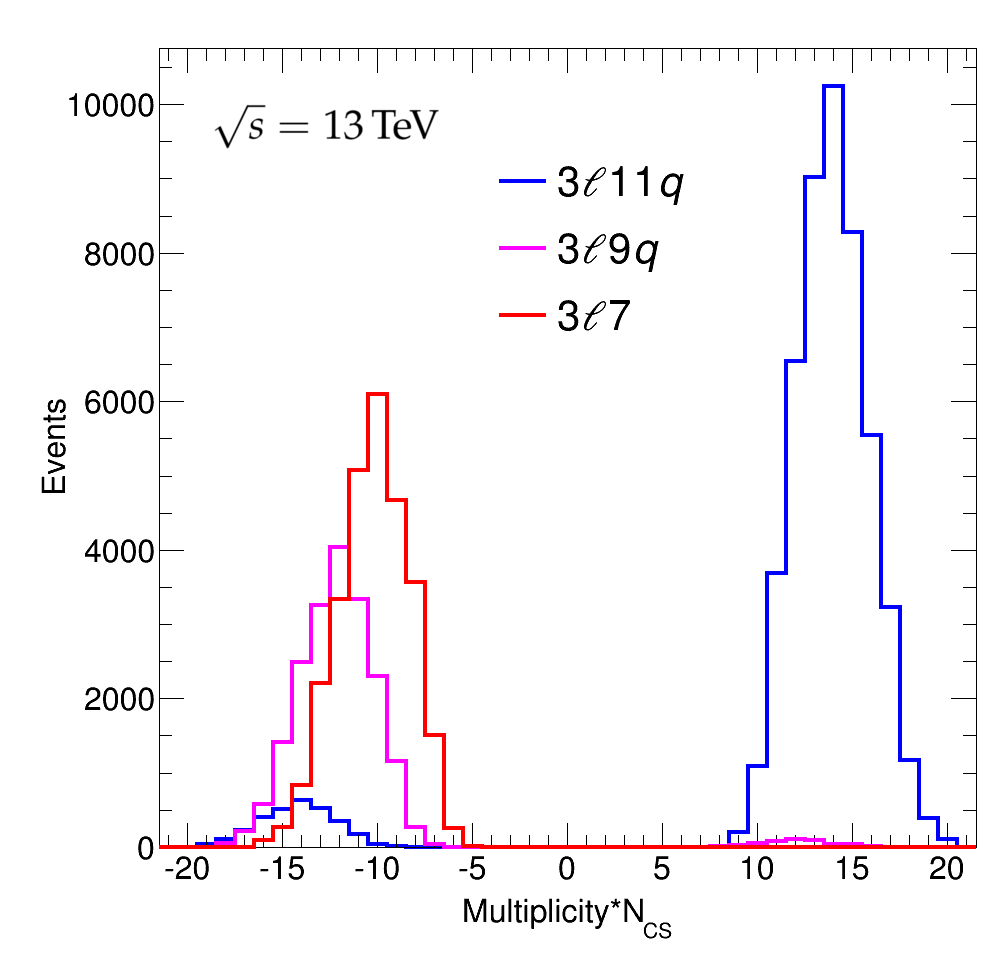}
    \caption{\label{fig:compNpartsSep} The distributions of multiplicities 
        of energetic ($p_T>20$~GeV) jets and leptons within a 
        nominal fiducial detector acceptance of $|\eta|<2.5$,
        after processing with PYTHIA, including top quark and $W$ decays, 
        as well as jet fragmentation and hadronization. 
        The multiplicities are plotted separately for $N_{CS}=\pm 1$ by multiplying
        them by $N_{CS}$, and are shown separately for each of the 
        possibilities for number of quark-antiquark cancellations (0, 1, and 2). 
        }
\end{figure}

The distributions of $N*N_{CS}$ at center-of-mass energy 13 TeV, 
after top quark and $W$ decays, for each of the number of potential incoming parton cancellations, is shown in Figure \ref{fig:compNpartsSep}.
This plot separates the distributions of $N$ by the sign of $N_{CS}$
(separating fermion from antifermion-type transitions), and corresponds
to Figure~4 (top left) of \cite{Ellis2016}.


\section{Using the Generator}

The BaryoGEN code is available as a public github repository \cite{BaryoGEN}. 
It is recommended that the general user download the most recent tag. The code has dependencies that are noted in the README included in the source code.
It is assumed that ROOT~\cite{ROOT}, LHAPDF~\cite{LHAPDF}, and CT10~\cite{CT10} 
have been correctly installed. 
In general, the program is run with the syntax:
\begin{verbatim}
BaryoGEN sqrtS threshold maxweight Nevents pNCS bCancel Filename
\end{verbatim}
If the number of arguments is not correct, the program will output this syntax reminder and terminate. 
The {\bf sqrtS} is the proton-proton center-of-mass energy in GeV.
The {\bf threshold} is the minimum energy in GeV required for a transition.
The parameter {\bf maxweight} must be set to a value that is greater than any 
possible probability given by the PDF set used,
and also should be changed any time the threshold energy is varied.
The {\bf pNCS} and {\bf bCancel} parameters will be described in the following paragraph,
and the {\bf Filename} is the name of the output files.

There are a few options for configuring the types of output events. 
The {\bf pNCS} parameter sets the probability of events to have $\Delta N_{CS} = +1$. 
The probability of getting $\Delta N_{CS} = +1$ is one minus this probability. 
In the results presented in this paper, we have set {\bf pNCS} to $0.5$ 
as mentioned at the end of Section~\ref{simresults}.
The generator also gives the ability to turn off the parton cancellations by setting 
{\bf bCancel} to 0 or to turn them on by setting it to 1. 
The configurations that correspond most closely to Ellis and Sakurai \cite{Ellis2016}
are with {\bf pNCS} set to $0.0$ or $1.0$ with parton cancellations on.

\begin{figure}
    \centering
    \includegraphics[scale=0.30]{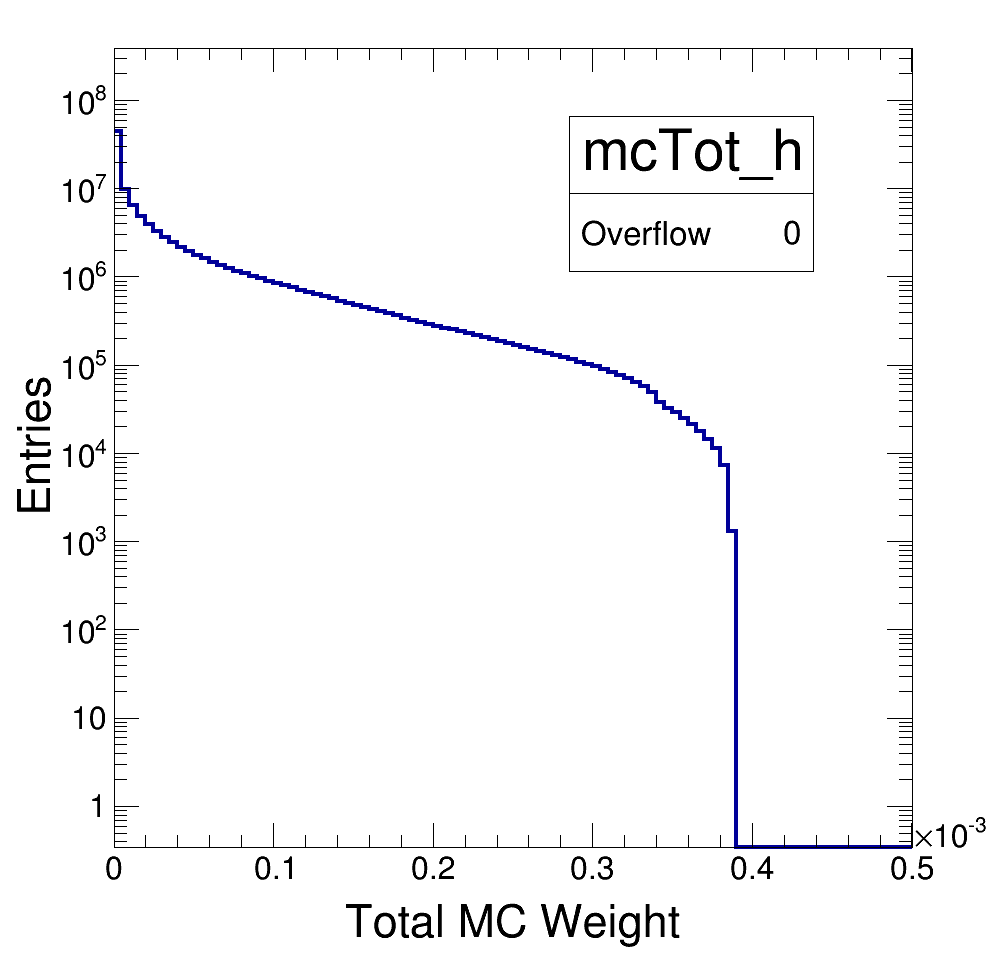}
    \caption{\label{fig:mcTot} An example of mcTot\_h (pp energy 13 TeV, $E_{Sph}=9$~TeV) that has a high enough maximum weight for the sample produced.}
\end{figure}

The following commands will build the executable and generate $10,000$ events. 
The generated output files will be called testRun.root and testRun.lhe. 
\begin{verbatim}
git clone https://github.com/cbravo135/BaryoGEN.git
cd BaryoGEN
make
./BaryoGEN 13000 9000 5e-4 10000 0.5 1 testRun
\end{verbatim}
These commands will output two files: testRun.root and testRun.lhe.
The LHE file can be used to further process the events with the user's software 
of choice for hadronization and fragmentation. The file testRun.root
contains histograms and an Ntuple of the generated events, 
to make it easy for the end user to analyze the generated events. 
The histogram mcTot\_h, as seen 
in Figure \ref{fig:mcTot}, is the most important histogram to check of these. 
If this histogram has any events in the overflow bin, the second 
argument given to the program must be increased so that the output is correctly weighted. 
This maximum weight parameter should only need to change if 
the proton-proton energy or the transition energy threshold is changed. 

\section{Conclusions}
We have a presented a new Monte Carlo generator, BaryoGEN, 
for simulating baryon and lepton-number violating sphaleron-like transitions in proton-proton collisions.
The BaryoGEN output uses LHE files to interface smoothly with PYTHIA 
for hadronization and fragmentation.
Kinematic distributions have been produced for comparison with the performance of other generators. 

It may be sufficient at current LHC energies and integrated luminosities to
search for sphaleron-like transitions by simply looking 
for excesses of high multiplicity, high transverse-energy events, since
the very high multiplicity and very high energy predicted for these transitions
leads to a very low level of QCD, top quark, and electroweak backgrounds.
Such searches are similar to those done for microscopic black holes 
\cite{CMSbh,ATLASbh}.
An ansatz that $N_{CS} = -1$ and $N_{CS} = +1$ transitions are equally likely 
is simple and seems sufficient in the absence of positive signals.

It may become necessary to reduce the background levels further in 
future searches for sphaleron-like transitions, in order to achieve maximum
sensitivity.
Such circumstances include a) greatly increased amounts of LHC data, 
b) higher proton-proton energies, 
c) searches for BSM sphaleron-like transitions with lower transition energies \cite{BSModel,SU3toy}, or d) if positive signals are found.
In such cases, an attractive, but more complex method would be to classify 
the events based on lepton content and missing transverse energy. 
Those channels containing charged leptons and/or large missing transverse energy
would have greatly reduced QCD background; and many of the channels would contain
same-sign charged leptons that are rare in the Standard Model.

BaryoGEN is an event generator that faciliates the study of a sphaleron-like 
class of $(B+L)$-violating transitions at LHC or future proton colliders. It is uncertain 
whether rates due to Standard Model sphaleron transitions are large 
enough to be observable, but such transitions could also arise from 
beyond Standard Model physics. In either case, these transitions would 
give rise to spectacular signatures, and could be the first direct 
evidence of baryon-number and lepton-number violation.

\appendix

\acknowledgments
Thanks to Graciela Gelmini for the original inspiration for this work, 
and David Saltzberg, Greg Landsberg, Steve Mrenna, John Ellis, Eric Cotner, 
Ji-Haeng Huh, and Doojin Kim for many useful discussions. Thanks also to 
Kazuki Sakurai for providing the cross-sections evaluated with various
PDF sets and energies.

\bibliographystyle{JHEP}
\bibliography{bib/sources}

%
%
%
%
%
%


\end{document}